# A Literature Review of Code Clone Analysis to Improve Software Maintenance Process

Md. Monzur Morshed* [1, 3], Md. Arifur Rahman[2], Salah Uddin Ahmed[1]
{m.monzur@gmail.com, monzur@scicon.org, monzur@tigerhats.org}
marahman@sce.carleton.ca, suahmed@aiub.edu
Department of Computer Science
American International University-Bangladesh[1], Carleton University-Canada[2], SCICON & TigerHATS-Bangladesh[3]

*Abstract*—**Software systems are getting more complex as the system grows where maintaining such system is a primary concern for the industry. Code clone is one of the factors making software maintenance more difficult. It is a process of replicating code blocks by copy-and-paste that is common in software development. In the beginning stage of the project, developers find it easy and time consuming though it has crucial drawbacks in the long run. There are two types of researchers where some researchers think clones lead to additional changes during maintenance phase, in later stage increase the overall maintenance effort. On the other hand, some researchers think that cloned codes are more stable than non cloned codes. In this study, we discussed Code Clones and different ideas, methods, clone detection tools, related research on code clone, case study.**

## I. INTRODUCTION

Code clone is one of the factors that make software maintenance more difficult [1]. A code clone is a code block in source files which is identical or similar to another code block. Code clones concept is of various reasons such as reusing code by 'code-and-paste' and others which make the source files very difficult to modify consistently. If faults found in one code block then the entire cloned blocks need modification and it becomes more difficult tasks to maintain if the system becomes big.

Recently, it is pointed out that maintenance phase is the most expensive one in the entire software development process. Many research studies have reported that large software companies spent a lot of cost to maintaining the existing systems [2]. Maintenance of software system is defined as modification of a software product after delivery to correct faults, to improve performance or other attributes, or to adapt the products to a modified environment [3].

Code clones are the source of heated debates among software maintenance researchers [4]. Many researchers consider clones to be harmful [5, 6, 7, 8, 9, 10, 11], due to the belief that inconsistent changes increase both maintenance effort and the likelihood of introducing defects. Yet, other researchers do not find empirical evidence of harm [12, 13], or even establish cloning as a valuable software engineering method to overcome language limitations or to specialize common parts of the code [14, 15, 16, 17]. It is not yet clear which of these two visions prevails, or whether the right vision depends on the software system at hand [18, 19, 20].

## II. RELATED RESEARCH

Y. Ueda et al. [23] developed a maintenance support environment based on code clone analysis called Gemini. Gemini delivers the source files to the code clone detector, and CCFinder [24] then represents the information of the detected code clones to the user through various GUIs.

Hotta et al. [32] showed a different approach on the impact of clones in software maintenance activities to measure the modification frequencies of the duplicated and non-duplicated code segments. According to their study, the presence of clones does not introduce extra difficulties in the maintenance phase.

M. Kim et al. [33] proposed a model of clone genealogy on clone evolution. According to their study, refactoring of clones may not always improve software quality based on the revisions of two medium sized Java systems during their study.

Krinke [34] found Type (I) code clones can be changed consistently during maintenance measured by Simian [35] (a code clone detector) and diff (a file comparison utility) on Java, C and C++. He also found that half of lifetime of clone groups consistently changing.

N. Bettenburg et al. [4] conducted an empirical study on three large open source software systems on the relation of inconsistent changes to code clones with software quality, at the level of official releases. In particular, they addressed the following four research questions:

(Q1) What are the characteristics of long-lived clone genealogies at the release level?
(Q2) What is the effect of inconsistent changes to code clones on code quality when measured at the release level?
(Q3) How does the effect of inconsistent changes to code clones at the release level compare to finer-grained levels?
(Q4) Which cloning patterns are observed at the release level?

According to R. Koschke et al. [31], there are still several open fundamental and terminological questions in software redundancy as follows:

(Q1) What are suitable definitions of similarity for which purpose?
(Q2) What other categorizations of clones make sense (e.g., semantics, origins, risks, etc.)?
(Q3) What is the statistical distribution of clone types?
(Q4) Are there correlations among orthogonal categories?
(Q5) Which strategies of removal and avoidance, risks of removal, potential damages, root causes, and other factors are associated with these categories?
(Q6) Can we create a theory of redundancy similar to normal forms in databases?

## III. CODE CLONE ANALYSIS

**a.) Definitions on Code Clone**

A clone relation is defined as an equivalence relation (i.e., reflexive, transitive, and symmetric relation) on code portions [21]. A clone relation holds between two code portions if (and only if) they are the same sequences. (Sequences are sometimes original character strings, strings without white spaces, sequences of token type, and transformed token sequences.) For a given clone relation, a pair of code portions is called clone pair if the clone relation holds between the portions. An equivalence class of clone relation is called clone class. That is, a clone class is a maximal set of code portions in which a clone relation holds between any pair of code portions [2].

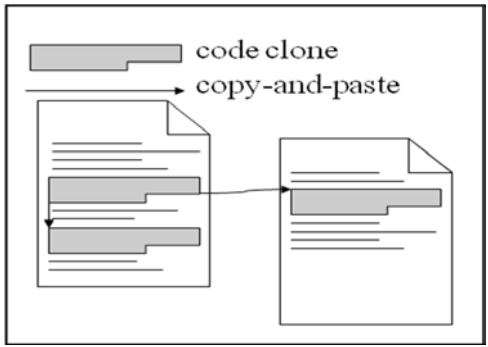

Figure 1: Code clone [22]

```
1 int sum = 0 ;
2
3 void foo ( Iteratoriter ){
4 for ( item = first ( iter ) ; has more ( iter ) ; item = next ( iter ) ) {
5 sum = sum + value ( item ) ;
6 }
7 }
8 int bar ( Iteratoriter ){
9 int sum = 0 ;
10 for ( item = first ( iter ) ; has more ( iter ) ; item = next ( iter ) ) {
11 sum = sum + value ( item ) ;
12 }
13 }
```

Figure 2: Example of Code Clones [31]

Code clone has no single or generic definition. Each researcher has own definition [22]. Usually clones are categorized in three types:

Type 1 clone: syntactical equivalence
Type 2 clone: parameterized syntactical equivalence
Type 3 clone: semantic equivalence

**b.) Clone Pair and Clone Set**

I.) Clone Pair: A pair of identical or similar code fragments
II.) Clone Set: A set of identical or similar fragments

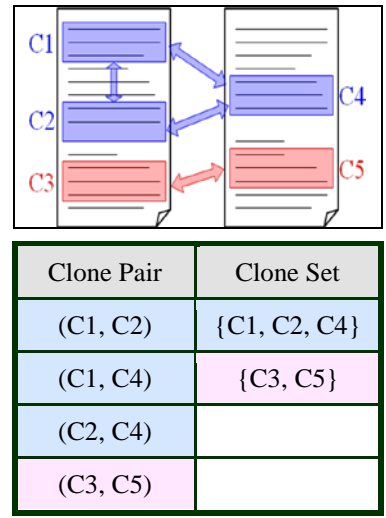

| Clone Pair | Clone Set |
|---|---|
| (C1, C2) | {C1, C2, C4} |
| (C1, C4) | {C3, C5} |
| (C2, C4) | |
| (C3, C5) | |

Figure 3: Clone Pair and Clone Set [22]

**c.) Advantages and Applications of Detecting Code Clones**

**Detects library candidates:** Code fragment proves its usability by coping and reusing multiple times in the system that can be incorporated in a library and announce its reuse potential officially [41, 42].

**Helps in program understanding:** It is possible to get an overall idea of other files containing similar copies of the fragment, if the functionality of a cloned fragment is understood. For example, when we have a piece of code managing memory we know that all files that contain a copy must implement a data structure with dynamically allocated space [43].

**Helps aspect mining research:** Detecting code clone is also necessary in aspect mining to detect cross-cutting concerns. The code of cross-cutting concerns is typically duplicated over the entire application that could be identified with clone detection tools [44, 45].

**Finds usage patterns:** The functional usage patterns of the cloned fragment can be discovered if all the cloned fragments of the same source fragments are detected [43].

**Detects malicious software:** To detect malicious software clone detection techniques can play a vital role. By comparing one malicious software to another, it is possible to find the evidence where parts of the one software system match parts of another [46].

**Detects plagiarism and copyright infringement:** Finding similar code may also useful in detecting plagiarism and copyright infringement [46, 47, 48].

**Helps software evolution research:** Clone detection techniques are successfully used in software evolution analysis by looking at the dynamic nature of different clones in different versions of a system [49, 50, 51, 52, 53].

**Helps in code compacting:** Clone detection techniques can be used for compact device by reducing the source code size [54, 55].

**d.) Drawbacks of Code Duplication**

Apart from beneficiary of code clones, it has severe impact on the quality, reusability and maintainability of a software system. The following are the list of some drawbacks of having cloned code in a system.

**Increased probability of bug propagation:** If a code segment contains a bug and that segment is reused by coping and pasting without or with minor adaptations, the bug of the original segment may remain in all the pasted segments in the system and therefore, the probability of bug propagation may increase significantly in the system [56, 57].

**Increased probability of introducing a new bug:** In many cases, only the structure of the duplicated fragment is reused with the developer's responsibility of adapting the code to the current need. This process can be error prone and may introduce new bugs in the system [58, 59].

**Increased probability of bad design:** Cloning may also introduce bad design, lack of good inheritance structure or abstraction. Consequently, it becomes difficult to reuse part of the implementation in future projects. It also badly impacts on the maintainability of the software [60].

**Increased difficulty in system improvement/modification:** Because of duplicated code in the system, one needs additional time and attention to understand the existing cloned implementation and concerns to be adapted, and therefore, it becomes difficult to add new functionalities in the system, or even to change existing ones [61, 62].

**Increased maintenance cost:** If a cloned code segment is found to be contained a bug, all of its similar counterparts should be investigated for correcting the bug in question as there is no guarantee that this bug has been already eliminated from other similar parts at the time of reusing or during maintenance. Moreover, When maintaining or enhancing a piece of code, duplication multiplies the work to be done [60, 62].

**Increased resource requirements:** Code duplication introduces higher growth rate of the system size. While system size may not be a big problem for some domains, others (e.g., telecommunication switch or compact devices) may require costly hardware upgrade with a software upgrade. Compilation times will increase if more code has to be translated which has a detrimental effect on the edit-compile-test cycle. The overall effect of cloning has been described by Johnson [61] as a form of software aging or "hardening of the arteries" where even small changes on the architectural level become very difficult to achieve in the actual code.

IV. CODE CLONE DETECTION TOOLS

Y. Higo et al. [2] developed a maintenance support environment based on code clone analysis called Gemini. T. Kamiya et al. [29] developed a code clone detection tool named CCFinder. Gemini delivers the source files to the code clone detector CCFinder, and then shows the information of the detected code clones to the user through various GUIs. Code Clone detection of CCFinder is a process in which the input is source files and the output is clone pairs [2].

V. METHODOLOGY PROPOSED BY RESEARCHERS

**Methodology I**

Y. Higo et al. [2] applied Gemini and CCFinder to several commercial software products in a case study where the users reported some problems as feedback and two problems were repeatedly reported and serious ones.

a.) The first problem shows itself in the case of 'copy-and-paste' reuse as the developers used the modified code portion. In the modification, developers also modified the user-defined identifiers in the code portion. In such case, the developers can subjectively identify the code clones even if they include some gaps among them where as CCFinder detects the clone as several short code clones separately. On the other hand, if the developers set a small value to the minimum length, then a lot of code clones are detected and practically the information is of no use [2]. Y. Ueda et al, [25] proposed the solution of this problem by referring to a certain set of gapped clones by representing renamed/modified code portions and gaps themselves on scatter plot.

b.) The Second problem shows itself if the developers detect code clones for refactoring [26], sometime semantically cohesive ones has more important meaning that maximal (just longest in local) ones although the formers may be shorter than the letters [2]. Their experiments found many clones that have not only primary logic statements but also the other coincidental clone statements before (and/or behind) them, since simple statements, such as assignment or variable declaration, tend to become clones coincidentally.

R. Komondoor et al. [27] and J. Krinke et al, [28] their approaches detect semantically cohesive code clones using program dependence graph (PDG) for procedure extraction. Since the cost to create PDG is very high and there are no examples exists based on their approaches to large scale software system. On the other hand, the clone detection process of CCFinder is very fast though lexical analysis can only be performed. So, the detected clones are not always semantically cohesive rather the clones are just maximal. As a result, the user of CCFinder has to extract semantically cohesive portions manually from the maximal. They solved this problem by a two-step approach in which they firstly detect maximal clones and secondly extract semantically cohesive ones from the results. Using this approach in real time they detect code clones that are easy to be reused [2].

**Methodology II**

S. Schulze et al. [30] showed two-staged approach to support code clone removal process. In the first stage, a detailed analysis of detected code clones is performed. In the second stage, the focus will be on how the results of stage 1 can be presented in order to guide an interactive refactoring/clone-removal process.

a.) The Analysis Stage encapsulates a detailed analysis process which is divided into one preprocessing and multiple post processing steps. The inputs for this stage are the detection results of existing code clone detection tools. They used CCFinder as code clone detector but according to them other tools can also be used. In the preprocessing, they merged code clones that have been detected to be similar to each other, to clone classes. After that, they classify these clone classes regarding the type of the cloned artifacts such as functions, loops etc. Finally, they investigated the clone classes which can be decomposed into smaller close classes. In case of finding, they divided the affected clone class [30].

b.) The result handling stage presents the gathered information of the analysis stage to the user. Therefore, it is important to provide the user with different views on clones as well as with abstract information on every clone's origin, based on that users can decide whether to remove them or not [30].

## VI. CODE CLONE CASE STUDIES BY RESEARCHERS

**Case Study I**

C. Kasper et al. [36] profiled the code cloning activity within a large software system Linux Operating System's Kernel that is widely used in industry. They profiled the code cloning activity to know more deeply how and why developers clone codes to enhance code clone detection process and code clone elimination strategies. They categorized different types of cloning activity based on the attributes such as location and size based on manual inspection of code clones found in the system. Their study produced a taxonomy of code cloning that will help other to examine code cloning. They used two methods to gather code clone information from the system. Firstly, they applied parameterized string matching that is implemented in CCFinder. Secondly, they applied metrics based code detection for which they used C/C++ to obtain raw metric information, as well as a set of Python scripts created to perform the code clone analysis.

**Case Study II**

E. Choi et al. [37] proposed a method to extract code clones for refactoring using clone metrics. They showed the usefulness of their proposed method based on a survey among developers in NEC Corporation. According to the feedback, it turned out their proposed method using combined clone metrics is effective method to extract code clones for refactoring [38]. Due to the time limitation, they conducted their previous study [37] on a single system. Therefore, their method may not generalize to other software systems. E. CHOI et al. [38] applied their proposed method [37] to open source software system and discussed their findings. In their study, they used two open source Java projects: Apache Ant [39] and JBoss [40] as their target systems. They also used CCFinder to detect code clone and 30 tokens as the minimum token length of a code clone as they followed the settings of previous study. According to their research results the conclusion of their case study says several clone sets are inappropriate for refactoring.

## VII. CONCLUSION & FUTURE WORK

In this paper, we conducted a literature review on code clone analysis to improve software maintenance process. We discussed several methods [2, 30] proposed by researchers to enhance code clone maintenance process. We also discussed advantages and drawbacks of code clone along with two case studies [36, 38] and the outcome of those case studies. As future work, we are planning to perform case studies of different open source projects and apply several methods that are proposed in several code clone studies.

REFERENCES

[1] M. Fowler, Refactoring: improving the design of existing code, Addison Wesley, 1999.

[2] Yoshiki Higo, Yasushi Ueda, Toshihro Kamiya, Shinji Kusumoto, Katsuro Inoue, "On Software Maintenance Process Improvement Based On Code Clone Analysis".

[3] Pigoski T. M, Maintenance, Encyclopedia of Software Engineering, 1, John Wiley & Sons, 1994.

[4] Nicolas Bettenburg, Weiyi Shang, Walid M. Ibrahim, Bram Adams, Ying Zou,Ahmed E. Hassan, "An empirical study on inconsistent changes to code clones at the release level", Journal Science of Computer Programming, pp. 1-17

[5] B.S. Baker, On finding duplication and near-duplication in large software systems, in: WCRE'95: Proceedings of the 2nd Working Conference on Reverse Engineering, IEEE Computer Society, 1995, pp. 86.

[6] I.D. Baxter, A. Yahin, L.M. de Moura, M. Sant'Anna, L. Bier, Clone detection using abstract syntax trees, in: ICSM'98: Proceedings of the 14th IEEE International Conference on Software Maintenance, IEEE Computer Society, 1998, pp. 368–377.

[7] R. Geiger, B. Fluri, H.C. Gall, M. Pinzger, Relation of code clones and change couplings, in: FASE'06: Proceedings of the 9th International Conference of Funtamental Approaches to Software Engineering, Springer, 2006, pp. 411–425.

[8] E. Juergens, F. Deissenboeck, B. Hummel, S. Wagner, Do code clones matter? in: ICSE'09: Proceedings of the 2009 IEEE 31st International Conference on Software Engineering, IEEE Computer Society, 2009, pp. 485–495.

[9] T. Kamiya, S. Kusumoto, K. Inoue, Ccfinder: a multilinguistic token-based code clone detection system for large scale source code, IEEE Transactions on Software Engineering 28 (7) (2002) 654–670.

[10] K. Kontogiannis, R. de Mori, E. Merlo, M. Galler, M. Bernstein, Pattern matching for clone and concept detection, Automated Software Engineering 3 (1–2) (1996) 77–108.

[11] A. Lozano, M. Wermelinger, Assessing the effect of clones on changeability, in: ICSM'08: Proceedings of the 24th IEEE International Conference on Software Maintenance, IEEE, 2008, pp. 227–236.

[12] F. Rahman, C. Bird, P.T. Devanbu, Clones: what is that smell? in: MSR'10: Proceedings of the 7th IEEE Working Conference on Mining Software Repositories, IEEE, 2010, pp. 72–81.

[13] S. Thummalapenta, L. Cerulo, L. Aversano, M.D. Penta, An empirical study on the maintenance of source code clone, Empirical Software Engineering 15 (1) (2010) 1–34.

[14] E. Duala-Ekoko, M.P. Robillard, Tracking code clones in evolving software, in: ICSE'07: Proceedings of the 29th International Conference on Software Engineering, IEEE Computer Society, 2007, pp. 158–167.

[15] C. Kapser, M.W. Godfrey, Cloning considered harmful considered harmful, in: WCRE'06: Proceedings of the 13th Working Conference on Reverse Engineering, IEEE Computer Society, 2006, pp. 19–28.

[16] C. Kapser, M.W. Godfrey, Supporting the analysis of clones in software systems: research articles, Journal of Software Maintenance and Evolution 18 (2) (2006) 61–82.

[17] M. Kim, V. Sazawal, D. Notkin, G. Murphy, An empirical study of code clone genealogies, in: ESEC/FSE-13: Proceedings of the 10th European Software Engineering Conference Held Jointly with 13th ACM SIGSOFT International Symposium on Foundations of Software Engineering, ACM, 2005, pp. 187–196.

[18] N. Göde, Evolution of type-1 clones, in: SCAM'09: Proceedings of the 2009 Ninth IEEE International Working Conference on Source Code Analysis and Manipulation, IEEE Computer Society, 2009, pp. 77–86.

[19] C.K. Roy, J.R. Cordy, Near-miss function clones in open source software: an empirical study, Journal of Software Maintenance and Evolution: Research and Practice 22 (3) (2010) 165–189.

[20] S. Thummalapenta, L. Cerulo, L. Aversano, M.D. Penta, An empirical study on the maintenance of source code clone, Empirical Software Engineering 15 (1) (2010) 1–34.

[21] T. Kamiya, S. Kusumoto, and K. Inoue, CCFinder: A multi-linguistic tokenbased code clone detection system for large scale source code IEEE Transactions on Software Engineering, 28(7):654-670, 2002.

[22] Katsuro Inoue, "Code Clone Analysis and Its Application", Software Engineering Lab, Osaka University

[23] Y. Ueda, T. Kamiya, S. Kusumoto, K. Inoue, Gemini: Maintenance Support Environment Based on Code Clone Analysis, 8th International Symposium on Software Metrics, pages 67-76, June 4-7, 2002.

[24] T. Kamiya, S. Kusumoto, and K. Inoue, CCFinder: A multi-linguistic tokenbased code clone detection system for large scale source code IEEE Transactions on Software Engineering, 28(7):654-670, 2002.

[25] Y. Ueda, T. Kamiya, S. Kusumoto, K. Inoue, On Detection of Gapped Code Clones using Gap Locations, 9th Asia-Pacific Software Engineering Conference, 2002

[26] M. Fowler, Refactoring: improving the design of existing code, Addison Wesley, 1999

[27] R. Komondoor and S. Horwitz, Using slicing to identify duplication in source code, In Proc. of the 8th International Symposium on Static Analysis, Paris, France, July 16-18, 2001

[28] Jens Krinke, Identifying Similar Code with Program Dependence Graphs, In Proc. of the 8th Working Conference on Reverse Engineering, 2001

[29] T. Kamiya, S. Kusumoto, and K. Inoue, CCFinder: A multi-linguistic tokenbased code clone detection system for large scale source code IEEE Transactions on Software Engineering, 28(7):654-670, 2002

[30] Sandro Schulze, Martin Kuhlemann, "Advanced Analysis for Code Clone Removal", University of Magdeburg, Germany

[31] Rainer Koschke, "Survey of Research on Software Clones", Dagstuhl Seminar Proceedings

[32] K. Hotta, Y. Sano, Y. Higo, S. Kusumoto, "Is Duplicate Code More Frequently Modified than Non-duplicate Code in Software Evolution? An Empirical Study on Open Source Software," Proc. EVOL/IWPSE, 2010, pp. 73–82

[33] M. Kim, V. Sazawal, D. Notkin, and G. C. Murphy, "An empirical study of code clone genealogies," Proc. ESEC-FSE, 2005, pp. 187–196

[34] J. Krinke, "A study of consistent and inconsistent changes to code clones," Proc of WCRE, 2007, pp. 170–178

[35] Simian similarity analyser. http://www.redhillconsulting.com.au/products/simian/

[36] Cory Kapser, Michael W. Godfrey, "Toward a Taxonomy of Clones in Source Code: A Case Study"

[37] E. Choi, N. Yoshida, T. Ishio, K. Inoue, and T. Sano, "Extracting code clones for refactoring using combinations of clone metrics". In Proc. of the IWSC 2011, pages 7-13, 2011

[38] Eunjong Choi, Norihiro Yoshida, Takashi Ishio, Katsuro Inoue, Tateki Sano, "Finding Code Clones for Refactoring with Clone Metrics: A Case Study of Open Source Software", Technical Report of IEICE

[39] The Apache Ant Project. http://ant.apache.org

[40] JBoss Application Server. http://www.jboss.org

[41] Neil Davey, Paul Barson, Simon Field, Ray J Frank. The Development of a Software Clone Detector. International Journal of Applied Software Technology, Vol. 1(3/4):219-236, 1995

[42] Elizabeth Burd and Malcolm Munro. Investigating the maintenance implications of the replication of code. In Proceedings of the 13th International Conference on Software Maintenance (ICSM'97), Bari, Italy, September 1997

[43] Matthias Rieger. Effective Clone Detection Without Language Barriers. Ph.D. Thesis, University of Bern, Switzerland, June 2005

[44] Magiel Bruntink. Aspect Mining using Clone Class Metrics. In Proceedings of the 1st Workshop on Aspect Reverse Engineering, 2004

[45] Magiel Bruntink, Arie van Deursen, Remco van Engelen, Tom Tourwe. On the Use of Clone Detection for Identifying Crosscutting Concern Code. Transactions on Software Engineering, Volume 31(10):804-818, October 2005

[46] Andrew Walenstein and Arun Lakhotia. The Software Similarity Problem in Malware Analysis. In Proceedings Dagstuhl Seminar 06301: Duplication, Redundancy, and Similarity in Software, 10 pp., Dagstuhl, Germany, July 2006

[47] Brenda Baker. On Finding Duplication and Near-Duplication in Large Software Systems. In Proceedings of the Second Working Conference on Reverse Engineering (WCRE'95), pp. 86-95, Toronto, Ontario, Canada, July 1995

[48] Toshihiro Kamiya, Shinji Kusumoto, Katsuro Inoue. CCFinder: A Multilinguistic Token-Based Code Clone Detection System for Large Scale Source Code. Transactions on Software Engineering, Vol. 28(7): 654- 670, July 2002

[49] Giuliano Antoniol, Gerardo Casazza, Massimiliano Di Penta, Ettore Merlo. Modeling Clones Evolution through Time Series. In Proceedings of the 17th IEEE International Conference on Software Maintenance (ICSM'01), pp. 273-280, Florence, Italy, November 2001

[50] G. Antoniol, U. Villano, E. Merlo, and M.D. Penta. Analyzing cloning evolution in the linux kernel. Information and Software Technology, 44 (13):755-765, 2002

[51] M.W. Godfrey, D. Svetinovic, and Q. Tu. Evolution, growth, and cloning in Linux: A case study. In CASCON workshop on Detecting duplicated and near duplicated structures in large software systems: Methods and applications, October 2000

[52] Ekwa Duala-Ekoko, Martin Robillard. Tracking Code Clones in Evolving Software. In Proceedings of the International Conference on Software Engineering (ICSE'07), pp.158-167, Minneapolis, Minnesota, USA, May 2007

[53] E. Merlo, M. Dagenais, P. Bachand, J.S. Sormani, S. Gradara, and G. Antoniol. Investigating large software system evolution: the linux kernel. In Proceedings of the 26th International Computer Software and Applications Conference (COMPSAC'02), pp. 421426, Oxford, England, August 2002

[54] W-K. Chen, B. Li, and R. Gupta. Code Compaction of Matching Single-Entry Multiple-Exit Regions. In Proceedings of the 10th Annual International Static Analysis Symposium ( SAS'03 ), pp. 401-417, San Diego, CA, USA, June 2003

[55] Saumya K. Debray, William Evans, Robert Muth, and Bjorn De Sutter. Compiler techniques for code compaction. ACM Transactions on Programming Languages and Systems (TOPLAS'00), Vol. 22(2):378-415, March 2000

[56] Zhenmin Li, Shan Lu, Suvda Myagmar, and Yuanyuan Zhou. CP-Miner: Finding Copy-Paste and Related Bugs in Large-Scale Software Code. In IEEE Transactions on Software Engineering, Vol. 32(3): 176-192, March 2006

[57] J Howard Johnson. Identifying Redundancy in Source Code Using Fingerprints. In Proceeding of the 1993 Conference of the Centre for Advanced Studies Conference (CASCON'93), pp. 171-183, Toronto, Canada, October 1993

[58] J Howard Johnson. Navigating the textual redundancy Web in legacy source. In Proceedings of the 1996 Conference of the Centre for Advanced Studies on Collaborative Research (CASCON'96), pp. 7-16, Toronto, Canada, October 1996

[59] Ira Baxter, Andrew Yahin, Leonardo Moura, Marcelo Sant Anna. Clone Detection Using Abstract Syntax Trees. In Proceedings of the 14th International Conference on Software Maintenance (ICSM'98), pp. 368-377, Bethesda, Maryland, November 1998

[60] Akito Monden, Daikai Nakae, Toshihiro Kamiya, Shin-ichi Sato, Ken-ichi Matsumoto. Software quality analysis by code clones in industrial legacy software. In Proceedings of 8th IEEE International Symposium on Software Metrics (METRICS'02), pp. 87-94, Ottawa, Canada, June 2002

[61] John Johnson. Substring Matching for Clone Detection and Change Tracking. In Proceedings of the 10th International Conference on Software Maintenance, pp. 120-126, Victoria, British Columbia, Canada, September 1994

[62] Jean Mayrand, Claude Leblanc, Ettore Merlo. Experiment on the Automatic Detection of Function Clones in a Software System Using Metrics. In Proceedings of the 12th International Conference on Software Maintenance (ICSM'96), pp. 244-253, Monterey, CA, USA, November 1996